# On Soft Mathematical Models of Subjective Time Acceleration with Age

**Vladimir Shiltsev,** *Fermi National Accelerator Laboratory, Batavia, Illinois 60510, USA*

RUNNING HEAD: [On time acceleration with age]

Correspondence address: Vladimir SHILTSEV, FNAL, PO Box 500, Batavia, Illinois 60510, USA; *shiltsev@fnal.gov* ; cell: +1 331-425-5277

**Abstract:** *It is a commonplace perception that speed of time subjectively experienced by humans significantly differs from chronological (objective) time and shows a great deal of variability. An often cited example is the phenomenon of the time acceleration with age – subjectively, the time passes faster as we get older. While the exact mechanisms behind it are not yet fully established, here we consider three "soft" (conceptual) mathematical models that might be applicable to the speeding time phenomenon: two proportionality theories widely discussed in the past and the original model that takes into account the novelty of experience effect. The latter is found the most plausible, as not only it satisfactorily describes the decadal subjective time acceleration, but also offers a reasonable explanation of the human life experience accumulation with age.*

## INTRODUCTION

So called "soft mathematical models" – understood as "… the art of getting relatively reliable conclusions from analysis of conceptual models" (Arnold, 1998) – are mathematical constructs allowing parametric predictions based on approximations of generally complex systems and their truncations to the most important dependencies and effects.  In the past "soft models" have been successfully applied to several phenomena in ecology, economics, sociology, and philology (Arnold, 2008; Elishakoff, 2019).  Here we review two most widely cited soft models of the subjective time acceleration with age, critically assess their assumptions and develop another model that can better approximate human experience.

The feeling that "The older we get, the faster time runs" is among the most common human perceptions - see voluminous observations of this phenomenon in psychology textbooks (James, 1890; Sternberg, Sternberg, & Mio, 2012) and experimental research literature (see, e.g., recent reviews in Hancock, 2002; Friedman and Janssen, 2010;  and Winkler et al., 2017), as well as popular books (Taylor, 2007; Hammond, 2013).  Obviously, the phenomenon in question is just one manifestation of human cognitive processing system (Sternberg, Sternberg, & Mio, 2012; Patterson et al., 2014), which by itself is very complex and consists of several major components, such as getting external inputs via sensory receptors (auditory, visual, tactile, etc), attention (a selection attempt to concentrate on sensory or mental surrounding events) and selection of the coded sensory messages before being transferred to the memory, the memory component comprising of short-term (working) and long-term memory (permanent one for concepts, mental images and cognitive maps), unconscious memory buffer store (Pang & Entlib, 2021), the processing mechanisms of storage (conceptualization, mapping and metal imagery) and retrieval (association and meaning-finding), as well as means to provide output response which in turn can influence external input. Though there is not yet a unified theory of cognition, the progress on cognitive architectures is very impressive with advanced models and theories successfully integrating regularities and assimilating prior research (Anderson, 1983; Evans, 2008; Ritter, Tehranchi, & Oury, 2019).

One of the contributing factors to the subjective acceleration of time is based on the physics of neural signal processing that leads to slowing the rate at which we process visual information , and this is what makes time 'speed up' as we grow older (Bejan, 2019). Of course, there are other sensory and cognitive phenomena contributing, too, and the most often offered explanations include ratio theories (see below), the age changing by number of memorable events, biological clock theories, attentional explanations, forward telescoping, difficulty of recall, and time pressure – see comprehensive review (Friedman & Janssen, 2010) and references therein.  Relevant experiments - see. e.g., (Friedman & Janssen, 2010; Winkler et al., 2017) - do not confirm predictions of several of these models, but indicate the reality of the age differences in the subjective speed of time when adults are asked to consider the last 10 years, while the effect is not present or very weak when they report on the last year or more recent time intervals. These and many other studies call for additional experimental and theoretical work to explain the determinants of this important aspect of adults' experience of time.

## PAST MODELS OF TIME PERCEPTION

One should start with a caveat that, as many other psychological phenomena, the acceleration of time is highly subjective and may vary not only with age, but also with cultural background, health, occupation, lifestyle, education, wealth, and many other factors. Time perception in younger children (Piaget, 2013) and now all so common end-of-life dementia are hard to study, evaluate and judge, and it can be quite different than the time perception for majority of population. Moreover, cognition and time passage are manifested at a variety of scales from tens of milliseconds, to minutes, to hours, days, weeks, months and years, while as pointed above, the speeding time effect is less prominent at intervals much shorter than a decade.

In mathematical terms, let $R$ be total real time and let $S$ be total subjective time. Small intervals of real time and subjective time are $dR$ and $dS$, respectively, where $dS$ is how long $dR$ feels. If we limit ourselves with just the long-term approximation (time perception over year to decadal scales), then the psychological time speeding would correspond to positive and decaying derivative $dS/dR\,(R_2) < dS/dR(R_1)$ if $R_2 > R_1$. Equivalently, one can define the relative acceleration $A(R,R_0)$ of the subjective time perception at age $R$ compared to that at any chosen reference age $R_0$, e.g., $R > R_0 = 20$ years:

$$A(R, R_0) = \frac{dS}{dR}(R_0) / \frac{dS}{dR}(R) > 1 \qquad . \qquad (1)$$

Historically, the first "soft mathematical model" was the so called *proportionality model* of the time perception (Janet, 1877, cited in James, 1886; James, 1890, and in Fraisse, 1963). It suggests that the subjective duration of a time interval decreases in inverse proportion of total real time $R$ (age):

$$dS_J(R) = C_J\, dR/R\ , \qquad (2)$$

where $C_J$ is a proportionality constant. For example, a single past year is subjectively 1/20 for a 20-year-old but 1/80 for a 80-year-old. Of course, the solution of this equation is a logarithm function:

$$S_J(R) = C_J \, ln(R) + const, \quad (3)$$

and the subjective time acceleration speed is simply:

$$A_J(R, R_0) = R/R_0, \qquad (4)$$

e.g., four times faster for $R$=80-years old individual than for the same individual at $R_0$=20 years: $A(80,20)= 80/20=4$.

Lemlich (1975) modified the proportionality model by noting that what is sensed is the subjective time, not the real clock time. Therefore, Eq.(2) has to be modified as :

$$dS_L(R) = C_L \, dR/S_L(R), \qquad (5)$$

where $C_L$ is a constant, so the resulting solution is:

$$S_L(R) = Q_L \, R^{1/2}, \qquad (6)$$

(naturally, $Q_L = (2 \, C_L)^{1/2}$) and the corresponding subjective time acceleration speed:

$$A_L(R, R_0) = (R/R_0)^{1/2}. \qquad (7)$$

In plain words - "the subjective duration of an interval of real time varies inversely with the square-root of the total real time (age)" (Lemlich, 1975: 235) and, correspondingly, the subjective time for a 80-years old passes twice as fast as for a 20-years old $A_L(80, 20) = (80/20)^{1/2}=2$.

There are a number of difficulties with these models, though. First of all, there are doubts in the postulated theoretical foundations because "we don't judge one day in the context of our whole lives" (Hammond, 2013: 158). Despite initial experimental confirmative indications (Lemlich, 1975; Walker, 1977) the results allowed other interpretations besides been questioned in (Friedman and Janssen, 2010) as based on trusting people's ability to accurately remember their experiences from long ago.

Also, the proportionality model has a singularity at $R=0$ (early in life) as $S_J(R)$ can get negative at small $R$. It was suggested in (Lemlich, 1975) to not consider too of small intervals, e.g., although a year is a finite interval of time, it is small in comparison with age and use of differentials in Eqs.(2) and (5) would still be justified. Nitardy (1943) suggested a modification of the proportional model by taking into account only the time elapsed since an individual has starting to recall memories, which he thought to be at the age of about 4 years. That would result in $A_J(R, R_0) = (R-4)/(R_0-4)$ and does not solve the singularity problem which would then move to $R=4$.

Easy to see that the simplest modification of Eq.(3) free of singularity would be:

$$S_J(R)=Q_J \, ln(1+R/\tau_J) \qquad\qquad (8)$$

with free parameters $Q_J$ and $\tau_J$. In this case, the subjective time $S_J(0)=0$ at birth and it approximately scales linearly with real time $S(R)\sim R$ at small $R << \tau_J$.

Weak point of the Lemlich's model was mentioned by the author himself – namely, that Eq.(6) gives "too rigid" of a prediction that, as a person ages, the subjective expected remaining life ($1-S/S_D$) will not be less than half his real expected remaining life. Indeed, if at death $R= R_D$ and $S= S_D$, then from Eq.(6) $S/S_D=(R/ R_D)^{1/2}$ and, correspondingly, $1-S/S_D =1-(R/ R_D)^{1/2} > \frac{1}{2} \, (1-R/R_D)$. It is hard to imagine *a priory* why it should be so, and, generally speaking, the model would need to be augmented by another parameter to avoid that.

All in all, there are arguments in favor of a search for a new model that would be free of the above mentioned deficiencies, at least in the mathematical sense.

## NEW MODEL OF SUBJECTIVE TIME ACCELERATION WITH AGE

Our model accounts for the effect of newness. As James (1890: 625) put it "…in youth we may have an absolutely new experience, subjective or objective, every hour of the day. Apprehension is vivid, retentiveness strong, and our recollections of that time, like those of a time spent in rapid and interesting travel, are of something intricate, multitudinous, and long drawn out. But as each passing year converts some of this experience into automatic routine which we hardly note at all, the days and the weeks smooth themselves out in recollection to contentless units, and the years grow hollow and collapse." The simplest mathematical construct to address this is to make the speed of accumulation of precepted subjective time inversely proportional to the weighted sum of the past gains $dS_S(R')$, with weight proportional to the time difference $(R-R')$ between the current age $R$ and time of the experience $R'$:

$$\frac{dS_S(R)}{dR} = \frac{C_S}{\int_0^R (R-R')\,dS_S(R')}, \qquad (9)$$

where the integration limits are from zero to current time (age) $R$. It is easy to see that the integral over $dS_S$ is equal to another integral over $dR'$, i.e., $\int (R-R')dS_S(R') = \int S_S(R')dR'$. Indeed, integration by parts results in $S_S(R')(R-R') + \int S_S(R')dR'$, and the first term is equal to zero if taken at $R'=R$ and at the $R'=0$ when $S_S(0)=0$. Such substitution leads to the solution that scales as:

$$S_S(R) = (2C_S\,ln(R))^{1/2}. \qquad (10)$$

Without loss of generality and with minimal effect on our following conclusions for ages $R \gg 1$, a slight modification of Eq.(10) allows to avoid unphysical singularity at birth ($R=0$) and make the initial subjective time (perception) to grow linearly with time:

$$S_S(R) = Q_S\,[ln(1+R^2/\tau_S^2)]^{1/2}, \qquad (11)$$

where $Q_S$ and $\tau_S$ are two parameters. Note, that this model predicts the subjective time acceleration with age scaling as:

$$A_S(R, R_0) \approx \left(\frac{R}{R_0}\right) \sqrt{\frac{\ln\left(1+\frac{R^2}{\tau_S^2}\right)}{\ln\left(1+\frac{R_0^2}{\tau_S^2}\right)}} \quad . \qquad (12)$$

### DISCUSSION

Despite very different underlying constructive foundations, the above models of the subjective time (which also may be referred to as accumulated experience or perception) $S_J(R)$, $S_L(R)$, and $S_S(R)$ – see Eqs.(4), (7) and (12), correspondingly - look quite similar at first glance, see Fig.1. They start fast at the beginning, at small $R$, and grow slower and slower with age $R$. To make a comparative evaluation, one should first demand that they all are properly normalized, so they result with the same individual's total accumulated subjective time at a certain age. At the same time, it is easy to see that the subjective time acceleration with age $A(R, R_0)$ does not depend on the choice of normalizing coefficients $Q_{J,L,S}$ – see Eq.(1). Table 1 presents the corresponding constants $Q_x$ and $\tau_x$ for the considered models if one normalizes the accumulated subjective time to that at 80 years of age, i.e., $S_x(80)$=1.0 ($x$=$J, L, S$). Already from there one can see qualitative rigidity of the Lemlich's model $L$ predicting that half of the life experience (elapsed subjective time) is lived by a certain age, namely by $R_{1/2}$=20 years for model $S_L(R)$. While we don't judge whether or not these precepted "half-life" ages are correct, it should be noted that due to an additional parameters $\tau_{S,J}$ models $S_S(R)$ and $S_J(R)$ are more flexible. For example, an all too common sentiment is that "… it's sometime said that human beings live two lives, one before the age of five and another one after" (Taylor, 2007: 11). By proper choice of $\tau_{J,S}$ both $S_J(R)$ and $S_S(R)$ can be made equal to 0.5 at $t_{1/2}$=5 years. Table 1 presents equations, parameters $Q_x$ and $\tau_x$ for the three soft mathematical models considered above, their corresponding "half-subjective-life" ages $R_{1/2}$ and ratios $A*(80,20)$ of the speeds of the subjective times at the ages of 80 years and 20 years (acceleration of subjective time).

Even bigger distinctions between predications of the models are observed in the decadal advances of life experience. Figure 2 presents the fractional values (in per cents of the total) of $\Delta S_{10}(R)=S_x(R)-S_x(R-10)$ for $R$=20, 30, 40, 50, 60, 70 and 80 years of age ($x=J, L, S$). One can see that while the decadal change from 10 to 20 years of age is about $\Delta S_{10}(R$ =0.13-0.16=13-16% for all four models, at the older age the decade from 70 to 80 years adds only 2% in $S_S(R)$ and more than 6% in the model $S_L(R)$. The latter seems unplausible and would result in small subjective acceleration of time with aging.

To evaluate the speeding of time with age one can calculate the ratio $A*(R)$ of the decadal life experience advance at the age of 20 to the same quantity at other ages:

$$A_x*(R)= [\ S_x(20) - S_x(10)\ ]\ /\ [\ S_x(R) - S_x(R-10)\ ]. \quad (13)$$

Figure 3 presents the results for all four models under test. If one could trust the common anecdotal reference that in comparing the older age (say, $R$=80) with younger one (say, $R$=20) time flies fast with "…one week as a day, one year as a month, one decade as a year or two", then foreseen value of $A*(80)$ should be between 5 and 10. As expected, the model closest to that is our $S_S(R)$ of Eq.(11) and the model most deviated is Lemlich's $S_L(R)$ of Eq.(6). It is to be noted, that while different models $S_x(R)$ may look similar in Fig.1, their corresponding observables, the perceived time acceleration with age $A_x*(R)$, are very different – as shown in Fig.3 and indicated in the last column of Table 1.

All the above models are essentially alternative mathematical expressions of the hypothesis that people experience fewer memorable events as they become older (therefore, the speeding of the subjective time) which can not be applied to any time scale except very long ones. Indeed, the timing of sub-second time scales involves automatic neurological processes (Lewis & Miall, 2003), whereas cognitive processes (attention and memory) of time intervals of seconds to minutes to be relearned ("performance feedback" - see, e.g., Ryan & Robey, 2002; Ryan & Fritz, 2007). In principle, one can expect that the timing of longer time scales, such as days, weeks, months, years, and decades would all similarly require different memory processes. Experimental studies so far report the time speeding phenomenon observed only when people are asked to compare the passage of the present time with the passage of past time, such as when they were half or a quarter of their present age. However, when participants are asked to judge the subjective experience of time of intervals, such as days, weeks, months, or years, then there are hardly any differences between age groups with the only exception of their experience of the last 10 years (see Winkler et al., 2017; Janssen, Naka, & Friedman, 2013; and extensive review in Friedman & Janssen, 2010). We believe that a much deeper insight into the feeling that time appears to pass faster as people become older requires a within-subjects approach rather than a between-subjects approach. Corresponding experiments should provide a life-long platform for a continuous look into how an individual perceives past and present memories, evaluate the functional form of the decay of the value or "freshness" of memorable events versus time passed or other events since, etc. The application of mathematical models like ours or similar ones could help us to assess human cognition structure and greatly contribute to the understanding of the time speeding phenomenon.

One can also remark that all the models $S_J(R)$, $S_L(R)$, and $S_S(R)$ are hardly testable at the early age $R \approx 0$. For example, they all imply that the accumulation of experiences starts at the very

birth - the claim that can not be easily substantiated. The proposed soft mathematical tune-ups at small $R$ - compare, e.g., Eqs.(3) and (8), Eqs.(10) and (11) – allow to avoid singularities in $S_J(R)$ and $S_S(R)$, but in general, there might be other ways and functional forms which potentially can better approximate experimental date if and when those will become available. In general, minor mathematic drawbacks of the proposed models may be justified by the difficulty to put a cognitive process into a formal context.

## CONCLUSIONS

We have considered three mathematical models of the subjective time acceleration, understood as slowing of human "mind clocks" with respect to real time clock with aging. The first two $S_J(R)$, and $S_L(R)$ represent variations of the proportional theory by Janet (1877) and Lemlich (1975) and summarize the accumulated life events perception as in Eqs.(3, 8) and (6), respectively. The proportional models have shaky, widely questioned logical foundation. We have come up with a new model $S_S(R)=Q[ln(1+R^2/\tau_S^2)]^{1/2}$, see Eq.(11), which takes into account the effect of fading out past memories with time. It is the most flexible out of all the models, is the closest to widely experienced fast speed of the perceived life at older age $R$ and predicts the subjective time acceleration with age scaling as $A(R)\sim R\cdot[ln(1+R^2)]^{1/2}$.

In this work, we did not aim at specific realism but at capturing the defining mechanisms behind the time speeding. We believe that the proposed soft mathematical model can be quite useful to define the experimental protocol of future experimental studies in the psychology of time perception and for the analysis of a large number of already reported observations of this phenomenon.

**Table 1**. Comparison of three soft mathematical models of the subjective time evolution – the one proposed here $S_S(R)$, $S_J(R)$ (Janet, 1877) and $S_L(R)$ (Lemlich, 1975): mathematical Eqs.(11, 8, 6), parameters $Q_x$ and $\tau_x$, "half-subjective-life" ages $R_{1/2}$ and ratios $A^*$ of the subjective time speeds at the ages of 80 years and 20 years (see text).

| Model | Equation | Half-life" age $R_{1/2}$, yrs | $\tau$, years | Q for $S_i(80)=1$ | Acceleration of subj. time $A^*=$ $\Delta S_{ri}(80)/\Delta S_{ri}(20$ |
|---|---|---|---|---|---|
| $S_S$, this work | $Q[\ln(1+R^2/\tau^2)]^{1/2}$ | 20 | 20 | 0.594 | 4.8 |
| | | 5 | 2 | 0.425 | 6.9 |
| $S_J$ (Janet, 1877) | $Q\,\ln(1+R/\tau)$ | 20 | 10 | 0.455 | 3.4 |
| | | 5 | 0.4 | 0.189 | 5.1 |
| $S_L$ (Lemlich, 1975) | $Q\,R^{1/2}$ | 20 | | 0.111 | 2.3 |

| Model | Equation | "Half-life" age $R_{1/2}$, yrs | $\tau$, years | $Q$ for $S(80)=1$ | Acceleration of subj. time $A^*=$ $\Delta S_{10}(80)/\Delta S_{10}(20)$ |
|---|---|---|---|---|---|
| $S_S$, this work | $Q\,[ln(1+R^2/\tau^2)]^{1/2}$ | 5 | 2 | 0.425 | 6.9 |
| | | 20 | 20 | 0.594 | 4.8 |
| $S_J$ (Janet, 1877) | $Q\,ln(1+R/\tau)$ | 5 | 0.4 | 0.189 | 5.1 |
| | | 20 | 10 | 0.455 | 3.4 |
| $S_L$ (Lemlich, 1975) | $Q\,R^{1/2}$ | 20 | | 0.111 | 2.3 |

# FIGURE CAPTIONS

**Fig. 1.** Three models of the accumulated subjective time $S_S(R)$ – solid line, $S_J(R)$ – dashed line, and $S_L(R)$ – dotted line - vs real time (age) $R$, normalized to full life experience at the age of 80 years. Parameters $\tau_S$=2 years and $\tau_J$=0.4 years are chosen to come up with half-life experience at $R_{1/2}$=5 years in $S_S(R)$ and $S_J(R)$, correspondingly (see text).

**Fig. 2.** The decadal advances of life experience $\Delta S_{10}(R)=S_x(R) - S_x(R-10)$ for models $S_S(R)$ – solid line, $S_J(R)$ – dashed line, and $S_L(R)$ – dotted line. The fractional values are calculated in per cents of the total for $R$=20, 30, 40, 50, 60, 70 and 80 years.

**Fig. 3.** Subjective speeding of time $A^*(R)$ in the models $S_S(R)$ – solid line, $S_J(R)$ – dashed line, and $S_L(R)$ – dotted line - for $R$=20, 30, 40, 50, 60, 70 and 80 years, normalized to that at the age of 20 years.

Fig.1

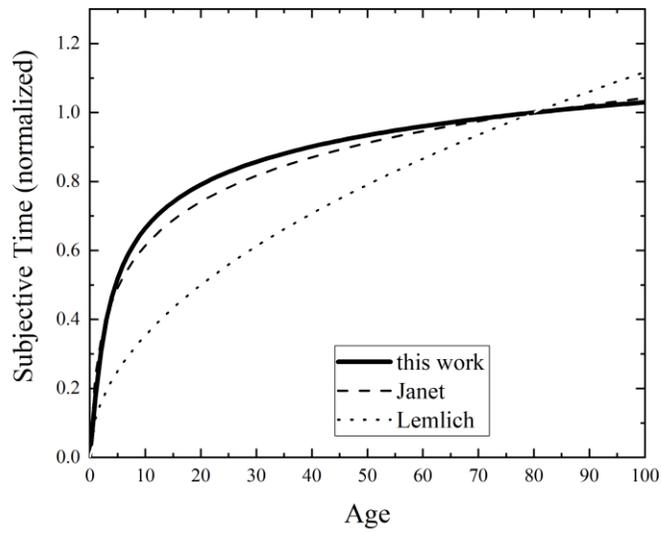

Fig.2

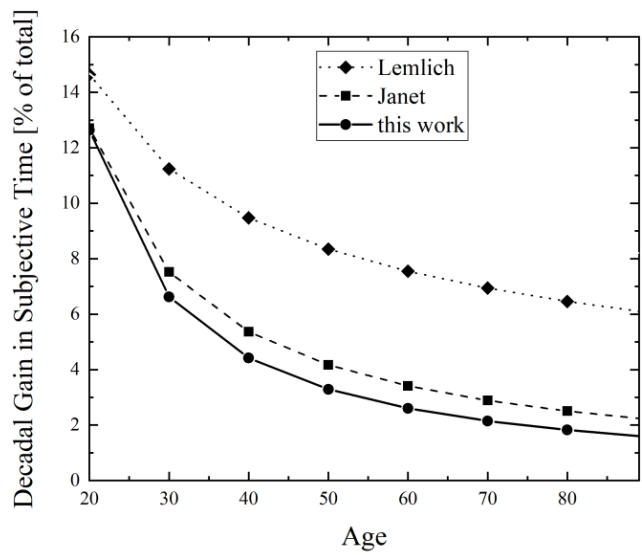

Fig.3

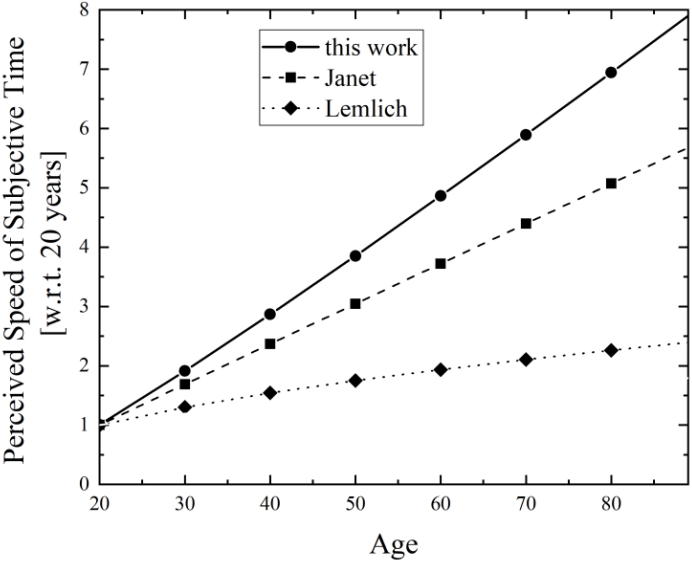